Barrie J Tonkinson February 2009
barrietonkinson@onetel.com

# The Behaviour of Clocks and Rods in Special and General Relativity

Barrie Tonkinson

ABSTRACT

While adhering to the formalism of Special and General Relativity, this paper considers the interpretation of clock rates and the rating of clocks in detail. We also pay particular attention to the crucial requirement of reciprocity between inertial frames. Our overriding concern is to bring out a distinction between clocks which run slow (slowly) in the everyday sense and those which record a smaller time interval between a specific pair of events - while running at the standard rate. The day by day application of relativistic formalism is not affected, but the underlying physics is changed.

# 1 Introduction

In a pedagogical paper of 1976, the distinguished physicist J S Bell gave his readers a simple question in Special Relativity. Bell had previously put the question to practicing theoreticians and experimentalists at CERN. Bell said "Of course many people who gave the wrong answer at first gave the right answer on reflection."

It is now over one hundred years since the publication of Einstein's seminal paper: *On the electrodynamics of moving bodies*, yet questions such as Bell's, which centred on the so called Lorentz contraction, still lead to controversy. There seems to be no diminution in discussions regarding clock rates and what effects there may be on measuring rods on account of their motion. Special Relativity (SR) is a bedrock of twentieth century physics which physicists use daily, so how can it be that contention continues to exist alongside this apparently satisfactory state of affairs? General Relativity, (GR), for most of the $20^{th}$ century, had rather less immediacy for many physicists and was less often the subject of heated debate. With the introduction of the Global positioning system (GPS) (and the Russian Glonass) there is a sense in which GR has became part of our daily lives; further, the GPS provides an ongoing experimental background - finding application in the better understanding of some fundamental questions.

At the beginning of the $21^{st}$ century it would not be unreasonable to expect that there should be a consensus view on what Special Relativity and General Relativity say about the behaviour of clocks and rods. While disagreements may emerge from time to time, perhaps most people would agree that the observed decay times of particles moving at high speed are explained by the slowing of their 'clocks' according to Special Relativity. In General Relativity, surely we are now comfortable with the corrections made to the satellite clocks of the Global Positioning System (both GR and SR corrections are made in the GPS).If there is a consensus, it is perhaps exemplified by J D Jackson in the (1999) third edition of his long established standard text, *Classical Electrodynamics,* "A moving clock runs more slowly than a stationary clock" and by H A Klein, in *The Science of Measurement* "…based on the General theory of Relativity, which says that physical events take longer - meaning that clocks run more slowly - in an intense gravitational field than in one less intense." Many physicists would argue that such views encapsulate what the theory says - and the theory is undoubtedly right. While the formalism is right, I take issue with the consensus view just outlined. My position is that clocks do not go slow (and rods do not contract). To establish this, we need a careful examination of what we mean by the rate of a clock.

## 2  The Rating of Clocks

We start with a standard clock, such as would be used by a rating authority; this immediately raises the question of how one might define a standard or good clock[1].We will not take up this difficult question in the discussion which follows; rather, it will be more appropriate to use the international standards for clocks - agreed worldwide by rating authorities[2]. The genesis of such standards is found in the longitude problem; this has relevance for us, so it will be useful to anchor our endeavour historically.

### 2.1  The Longitude Problem

Solving the longitude problem was to do with making a clock which would sustain a sufficiently consistent rate at sea, so that an accurate time was at hand for taking astronomical sights. When accurate chronometers became available in the latter part of the eighteenth century[3] the practice evolved for a ship's chronometer to be checked or rated. On return to port the chronometer would be taken to a rating authority. Here it would be placed alongside a clock maintaining standard time, in controlled conditions. At the end of the rating period a certificate would be issued for the clock under test - giving the rate of loss or gain in indicated time (say in fractions of a second per day). It is precisely this way of checking

---

[1] See, for example, Kilmister and Tonkinson
[2] Bureau International de l'Heure (BIH) Paris, U.S. Naval Observatory, National Physical Laboratory UK, etc.
[3] Betts J describes how Harrison at last won the great longitude prize for his amazing clock in 1773

clocks which we want to take guidance from for our initial considerations in relativity theory.

## 2.2 Classical Localised Rating

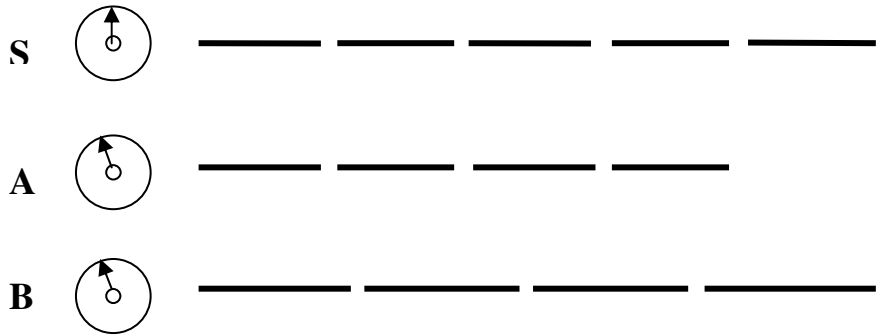

**Figure1**. Depicting the beat intervals of clocks **A** and **B** undergoing a rating check alongside the standard clock **S**. Only the minute hands are shown. The standard clock has run for 12hrs. Clock **A** was stopped after 11hr 55min, but kept perfect time up to that point. Clock **B** ran for the full 12hr period, but only recorded 11hr55min, it is said to run slow (slowly) in the everyday sense.

We consider a standard clock **S** in a controlled environment - as maintained by a rating authority. Two other clocks, **A** and **B**, which are to undergo rating checks, are placed alongside the standard clock. The three clocks are zeroed and then set running at precisely the same instant, which we call event 1. After 11hrs and 55mins, as recorded by **S**, clock **A** is stopped, this is designated event 2. At the 12hr point (event 3), clock **B** is stopped. Paper trace printouts (Figure 1), taken from each clock, are now compared. It is found that the ticks of **S** and **A** are in perfect synchrony up to event 2 when **A** was stopped; at this event **S** and **A** read 11hrs 55mins. On the other hand, when the printouts of **S** and **B** are compared, at the 12 hour point, event 3, according to **S,** it is found that the ticks of **B** fall slightly later than the standard ticks of **S.** The result is that, at event 3, clock **B** records only 11hrs 55mins.

Clock **B** is said to run slow (slowly); this is the normal or everyday sense of being slow. The certificate issued by the rating authority would state that **B** loses 5 minutes in every standard period of 12 hours. Provided that the rate was steady (linear, not unpredictably variable), then a correction could be applied to its readings to give standard time. In contradistinction, clock **A** kept perfect time up to event 2 when it was purposely stopped. **A** has a zero rate and it therefore displayed standard time while it was running.

We now compare the number of ticks produced by clock **A,** up to event 2, with the number of ticks produced by clock **B** up to event 3 and find that they are identical. This is because both clocks have the same reading of 1155 at their respective final events (if we think of mechanical clocks, with an escapement action, a given reading corresponds to so many operations of the escapement).

Let us now consider the total situation at event 3. While we know that **A** was stopped 5 minutes ago, **A** and **B** have identical readings at this event. What we can say is that, although both clocks record the same time (1155), **A** ran for less time than **B** - as judged by the standard clock. Most important, the presence of the standard clock is essential if we are to make such statements. The distinction between clocks which run slow in the normal sense and those which record less time, relative to precisely specified event sets, will be crucial in our further discussion. In these initial considerations both the standard clock **S** and clock **A** are in the same inertial frame (a non-relativistic case) and, necessarily, event sets specific to each clock have been used. The power of our approach will become clear when, relativistically, time intervals in two different inertial frames, employing a single event set, are compared.

So far we have only considered a localised means of comparing the rates of clocks; but it contains the essence of what is needed for application to relativity theory. We have signalled the distinction between those clocks which go slow in the normal sense and those which run at the standard rate for less time according to a standard clock. Moving to rather more global considerations, the first difficulty to overcome in clock comparison is that of spatial separation; for this it is necessary to assign time at points over an extended region.

### 2.3   Rating Distant Clocks

Figure 2 depicts an inertial reference frame with a single spatial axis and a time axis. A pulsed sequence of left and right going light signals is originated at the mid point between a set of mirrors. Reflected light oscillates between the mirrors, thus forming a field of light clocks,[4] as for example indicated by events 1 and 2

Repeated reflections set up event sequences which we call ticks - forming an extended set of light clocks in F. Each mirror both reflects and allows transmission, so that signals pass through the first pair of mirrors to be reflected and transmitted at events 3 and 4 and so on. The temporal extension of the mirrors has been only partially represented for clarity. Lower pulse recurrence rates at the more distant clocks are easily allowed for and a standard frame time is thereby established

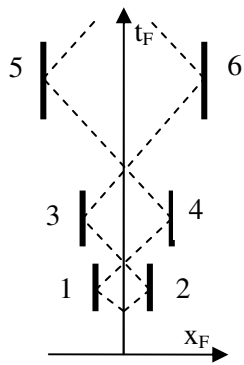

**Figure 2.** A single inertial reference frame employing a pulsed sequence of left and right going light signals forming a field of light clocks.

We have viewed the light signals as elements of a spatially extended light clock structure; of course at each mirror the tick sequences form local light clocks. Ordinary clocks can now be positioned at the various mirror locations and synchronised with the light clock ticks. The extended light structure provides the two elements needed for clock synchronisation: the establishment of an initial or zero datum (as say for Coordinated Universal Time [UTC]) and a means of ensuring that all local clock rates are the same across the frame of reference.
We use Einsteinian synchrony, so that for a symmetric event pair such as 3 and 4, we write $_3t_F = {_4t_F}$, meaning: the time at event 3 in frame F is equal to the time at event 4 in F.

---

[4] See, for example, Marder, L.

The employment of synchrony according to Einstein may be of some concern and raises the question of (so called) conventionality. Standard (Einsteinian) synchrony is used in this paper, chiefly because any other (non-standard) synchrony implies that the velocity of light is not equal in all directions. This argument is easily met by the conventionalists on the grounds that it is only the round trip speed of light which can be measured. We should not therefore pretend to knowledge about the speed of light in a single direction - which does not admit of measurement[5] On the other hand one can hold that, in spite of the impossibility of measuring the one way velocity, there may be a fact of the matter. This I take to be that the velocity of light is a physical constant of nature (as indeed brilliantly found by Maxwell and, earlier, Weber and Kohlrausch[6]) - embodying the implication that the velocity of light is isotropic.

So far we have addressed the first difficulty in clock comparison (that of spatially separated clocks) by providing a standard time at all points in a single inertial frame. We next consider the second difficulty - that of comparing clocks in different translating frames.

## 2.4   Clock Comparison across Relatively Translating frames

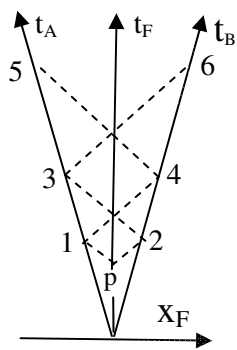

**Figure 3**. Inertial frames A and B translating left and right, each with uniform speed u in F. The same event set is used as in Figure 2. A passes through events 1, 3 etc., while B passes through the even numbered events.

---

[5] Chang makes a suggestion to measure the one way velocity of light, while Flidreynski, and Nowicki counter this view.

[6] See the Maxwell papers, XXXVI. On a Method of Making a Direct Comparison of Electrostatic and Electromagnetic Force; with a Note on the Electromagnetic Theory of Light. This paper makes reference to Weber and Kohlrausch.

We retain our reference frame F and introduce inertial frames A and B translating left and right - each with uniform speed u in F (fig.3.) The event set is the same as before. A now passes through events 1, 3 etc. while B passes through the even numbered events. We can relate the time in A to the time in B and relate these times to the time in F[7]. Light signals, originating in F at event p arrive in A and B at events 1 and 2 respectively. Signal transit times depend on frame speed and the speed of light; thus we apply a factor k = k(u,c) to a time interval in F to obtain a related interval in A or B[8] (see also relativistic doppler shift, p.12). Equation (1) reads: the time at event 1 in A = the time at event 2 in B = k multiplied by the time at event p in F.

$$_1t_A = {_2t_B} = k_p t_F \tag{1}$$

What is important for our immediate purpose is that the time in frame A at event 1 is equal to the time in frame B at event 2. We call such event pairs in relatively translating frames *Reciprocal Events*. Following our discussion for the fixed frame alone, the signals originated in F now form light clocks in each moving frame. If indeed it is the case that good clocks in A and B are synchronised at event zero, then they will record the same elapsed time at events 1 and 2 respectively. On reception of their signals at events 1 and 2, A and B can exchange radio signals and mutually confirm that their clocks read the same (similarly for events 3 and 4 etc.) Importantly we see that *clocks in the uniformly translating frames A and B run at the same rate*.

All of our considerations are based on time; whether clocks are in close proximity, spatially separated, or uniformly translating, we have seen that rate comparisons can be made using the same light signal network. Clock readings can be directly compared by the exchange of radio signals between translating frames.

We now introduce some current experimental considerations. Returning to frames A and B, they can be thought of as two free falling satellites - moving away from each other in relative translation. Frame F and associated geometrical constructions are no longer needed. One satellite clock can be regarded as master and send initial synchronising pulses to the other; mutually exchanged signals will be equally relativistically doppler shifted and can be used to adjust synchrony and monitor clock rates. Any rate errors will be detected and can be corrected. This situation relates to the ongoing monitoring of satellite clocks in the Global Positioning System (GPS). A brief note on the (GPS) is now included.

---

[7] Milne's first problem in time keeping, *Kinematic Relativity*, p16.
[8] Kilmister, C.W. [1970], p.17.

The GPS employs 24 orbiting satellites which need to be considered by using Special and General Relativity. A universal time is a requirement of the system, but clearly the satellites are not all in the same inertial frame. In order to construct an operational system, a single agreed time is introduced. This is made possible by assuming an underlying inertial frame in which the whole global system is immersed[9]. Pulsed time signals from the satellites, received at monitoring stations on the earth, will be shifted in frequency in accordance with SR, GR and doppler effects. Data can now be passed from a monitoring station to a satellite in order to bring its clock into conformity with the rate required by the assumed inertial frame. In this way a universal time is constructed for the system. The exchange of signals in our example is thus experimentally supported by the daily monitoring of GPS satellite clocks. As noted, equal relativistic doppler shifts in A and B will confirm equal clock rates (Gravitational effects are considered later in this paper). Thus, in contradistinction to a common view, there can be no question that a clock might in some sense be slowed on account of its motion. As we have emphasised, the rate of a clock is only defined relative to a standard.

### 3  Reciprocity between Inertial Frames

The key consideration in any discussion of relatively translating frames is reciprocity, this is the need for the absolute indistinguishability of the reference frames employed. We first

---

[9] Ashby, N. and Allan, D.W

recall Einstein's Principle of Relativity - which relates to the notion of Reciprocity.

Quoting from Einstein's 1905 paper: 'Let us take a system of co-ordinates in which the equations of Newtonian mechanics hold good. In order to render our presentation more precise and to distinguish this system of co-ordinates verbally from others which will be introduced hereafter, we call it the stationary system[10]'. As Einstein makes clear, this is only a verbal distinction; however it was a choice of words about which there has been a century of misunderstanding  We should read Einstein as follows: take any pair of uniformly translating inertial frames (represented by coordinate systems) and then, quite arbitrarily, nominate one, say K, as stationary (only a verbal distinction). The other frame, K', (for the time being we call this the travelling frame) will be viewed from K as it travels to a distant event in K. We see that Einstein's verbal distinction was acceptably descriptive of the temporary situation envisaged, but there was to be no physical difference between K and K'. Rather than such terms as resting or stationary frame, it would be better to introduce the term *fiducial frame.* This more clearly indicates an arbitrary choice of frame against which measurements can, for the time being, be made.

Returning to the notion of reciprocity between two inertial frames. Torretti[11] gives the following definition 'If the inertial frame F' moves in the inertial frame F with the velocity v, then F moves in F' with velocity -v.' This equality of speeds approach motivates a verbal description: *In whatever way we consider one frame of reference, it must be possible to consider the other frame in exactly the same way*. Reciprocity, while not Einstein's Principle of Relativity as such (which, rather, is specific about Laws), continues the spirit of the Relativity Principle. The need to obey the reciprocity condition at all times is an extremely stringent requirement. Thus, whenever a frame has been nominated as fiducial, it must be possible, at any instant, to switch the choice to the other frame and obtain a consistent outcome. So far we have emphasised the equality of clock rates in all inertial frames. This is a vital aspect of the reciprocity condition and of course is in stark contrast to any notion that a clock in a particular inertial frame goes slow (in the normally accepted sense)

---

[10] Kilmister, C. W. [1970], p188, translation of the 1905 paper.
[11] Torretti, R. p.79.

## 4 The Doppler Shift

Classically, a medium provides an absolute frame of reference in which it is possible to identify the motion of the wave source and/or observer. Importantly, this asymmetry enables us to distinguish the two frames, hence reciprocity is lost. Thus it is not merely the case that a medium is, so to say, superfluous[12], but, more strongly, that it is not possible to admit a medium into Relativity Theory. We will briefly examine the classical case, since it highlights the emergence of the reciprocal equality of two relativistic frames; however, we first need to derive the Relativistic Doppler Shift.

### 4.1 The Relativistic Doppler Shift

Nomenclature, $_{cd}t_A$ reads: the time between events c and d in frame A. We will also use shorthand: $_d t_A$, meaning; the time interval between an agreed initial or zero event and event d in frame A. Equation (2) below reads: The time interval between the initial event and event 2 in frame B equals k multiplied by the time interval between the initial event and the event *a* in the frame A .

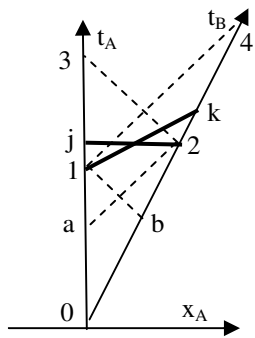

**Figure 4.** A more familiar representation of Figure 3, dispensing with frame F. Standard measuring rods extend between the frames; a and b are reciprocal events.

---

[12] Einstein used this term in the second paragraph of his 1905 paper. Translation in Kilmister, C. W. [1970], p.188

Referring to Fig.4, we calculate the factor $k = k(v, c)$ for the relativistic doppler shift. We have:

$$_2t_B = k\,_at_A \tag{2}$$

$$_3t_A = k\,_2t_B = k^2\,_at_A . \tag{3}$$

Using standard synchrony,

$$_2t_A = \tfrac{1}{2}[_at_A + _3t_A] = \tfrac{1}{2}\,_at_A[1+k^2] \tag{4}$$

This is the travel time of B, measured in frame A, when intercepted by the light signal (from event a) at the event 2. The travel time of the light signal is therefore

$$_2t_A - _at_A = \tfrac{1}{2}\,_at_A[k^2 - 1] . \tag{5}$$

The travel time of B and the travel time of the light signal are in the inverse ratio of their respective velocities. Where c is the speed of light, we have:

$$[k^2 - 1]/[1 + k^2] = v/c \tag{6}$$

$$k^2(c-v) = (c+v) \tag{7}$$

$$k = \sqrt{\frac{1+v/c}{1-v/c}} = [1+v/c][1-v^2/c^2]^{-\tfrac{1}{2}} \tag{8}$$

This is the relativistic doppler shift. The first bracket gives the classical doppler shift factor and the second is the Lorentz factor γ.

## 4.2 Classical Doppler - Relativistic emergence of Reciprocity

We can now consider the classical case. A source in frame A is at rest in a medium which supports wave velocity u. An observer (frame B) moves away from the source, and thus through the medium, with speed v. Let the source emit a pulsed waveform, the first pulse being at $_0t_A$ (the starting or zero time of the clock in A at event zero); the second pulse is emitted at $_Tt_A$ (the pulse period) The observer, on passing the source receives the first pulse at $_0t_B$ and sets her time to $_0t_B = {_0t_A}$ (since this is a classical case, both frame times are the same).

The second pulse is received in B at $_2t_A = {_2t_B}$. (9)

We equate the distance that frame B has travelled in A with the distance the second pulse has travelled when it intercepts B.

$$v_2t_A = u(_2t_A - {_Tt_A}) \quad (10)$$

So that the relationship between the period at transmission and that at reception is:

$$_Tt_A = (1 - v/u)\,_2t_A \quad (11)$$

Now we consider the source A moving through a medium in which the observer B is stationary. A emits the first pulse on crossing B at time $_0t_A = {_0t_B}$ (the time of reception in B). A emits the second pulse at $_Tt_B = {_Tt_A}$, B receives this pulse at $_2t_B$, thus,

$$v_Tt_B = u(_2t_B - {_Tt_B}), \quad (12)$$

$$_2t_B = (1 + v/u)\cdot{_Tt_B} \quad (13)$$

By transposing (1-v/u) in (11) and exchanging frame suffixes, we get (13), but only to first order, thus it is possible to distinguish the two frames. In the classical case there are extreme cases of the inequivalence of the reference frames. For example, it is possible to observe waves of the shock type (zero period) if the source has wave velocity through the medium; on the other hand, the period becomes infinite if the observer recedes at wave velocity.

In the classical treatment the medium provides a frame in which the source and observer are considered to be stationary in turn. Relativistically, no such medium exists, but if we follow the classical approach for the relativistic case, analogous choices can be made for the source and observer. Relativistically, quite arbitrarily, the source/observer can be nominated to be "at rest" (see below). To get the relativistic equations from the classical equations, we need to transform on the time and replace the original wave velocity u by the invariant light velocity c. Thus, transforming on the time and setting u = c, the frames are seen to be equivalent and reciprocity emerges. Equation (9) then becomes:

$$_2t_A = \gamma \,_2t_B, \qquad (14)$$

and the relativistic form of (11) is:

$$_Tt_A = \gamma(1 - v/c) \,_2t_B \qquad (15)$$

Similar considerations for the moving source (Eqn. 13) give:

$$_2t_B = \gamma(1 + v/c) \cdot _Tt_A \qquad (16)$$

On multiplying out by $\gamma$, (15) and (16) are seen to be identical and the classical asymmetry is removed. Indistinguishability in the relativistic case is a no medium characteristic. "Moving source" and "moving observer" become terms which only have a classical reference and are no longer applicable relativistically. Relative velocity now does all the work of the two classical terms.

The transverse doppler effect is a special case which occurs when the velocity vectors of the two inertial frames are orthogonal to the signal propagation axis. A velocity of approach now momentarily changes to a velocity of recession and we are left with just the Lorenz factor.

## 5    Interpreting the Minkowski Metric

We have been using events, but now consider such events as elements of a Minkowski spacetime in order to establish metrical relations. Again we take two frames, A and B. Where s is the invariant spacetime interval, the Minkowski metric is

$$\Delta s^2 = \Delta t_A^2 - \Delta x_A^2 = \Delta t_B^2 - \Delta x_B^2 \ . \tag{17}$$

We see immediately that the frame which finds a smaller spatial interval between a given pair of events will find a smaller time interval and vice versa. The Minkowski metric carries no information about clock rates[13]; as we have emphasised, clock rates are defined by comparison with a standard. A Minkowski time interval is the coordinate time interval, according to standard clocks, between a specific pair of events in a given inertial frame.

A Minkowski space interval is the radar distance, derived from the out and return time interval to a distant event in a given inertial frame. Identically, it is the distance found by a standard measuring rod, between a pair of simultaneous events. See under Measuring Rods below.

Let us take it that frame B finds a smaller time interval, between a given pair of events, than does frame A. *We say that, with the clocks in each frame running at the same rate between the given pair of events, clock B has run for a smaller amount of time than has clock A* (see p5 under Classical Localised Rating, but now we have a single event set).

### 5.1    MEASURING RODS

The description of measuring rods (definition of length) is closely linked to time in relativity theory. In current metrology practice, length is defined in terms of time. The time standard for length replaced the method of counting wavelengths in 1983[14]; both of these length standards are used in daily practice. Importantly, any standard physical measuring rod (say, made of invar), in any inertial frame, will conform precisely to current time or wavelength determinations in that frame.

We briefly recall the Lorentz transformations: t = γ(t'+vx'/c$^2$), that is, when the distance in the dashed frame is zero, the respective frame times are simple γ multiples. Again, x = γ(x'+ vt'), that is, when the time in the dashed frame is zero the distances are simple γ multiples. We see that the transformations for time and distance are similar - as expected, since we use time to find distance.

---

[13] More usual, one would say, is quite the opposite view, e.g. MTW, p.1054.
[14] Petley, B.W. [1983].

In Fig.4 (p11), measuring rods between events j and 2 and between k and 1 are depicted. The light paths *a*,2,3, and *b*,1,4, being initiated at the reciprocal events *a* and *b* (i.e. after the same amount of time has elapsed in each frame ) are mirror images and thus of equal length. Equivalently, the time interval between *a* and 3 in A is equal to the time interval between *b* and 4 in B. These signal paths can be thought of as radar measurements giving the length of the measuring rod j,2 in A and the length of rod k,1 in B. Thus, employing current metrological practice, the rods in each frame are found to be of equal length. We can now easily make the usual length (distance) comparisons, for specific event pairs, between the two frames.

Knowing the time of an event immediately gives its distance, but we need to be careful to use times reckoned from the relevant reciprocal events. For example, it is only necessary to ensure $_a t_A = {_b t_B} = 0$ to give correct light path travel times and hence distances.

## 6 Examples in Special Relativity

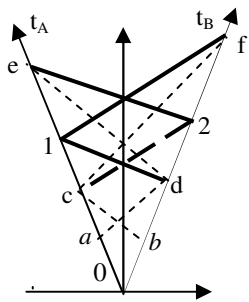

**Figure 5.** Inertial frame B translates to the right in A, while A translates to the left in B. Various measuring rods are shown.

We now take a standard example and refer to Fig.5. A symmetrical diagram has been reintroduced in order to visualise the equivalence of the two frames and various event sets more easily. Referring to Fig.5, there will be an event (0) which we call the initial event. B translates to the right in A; equivalently, A translates to the left in B While B is observed from A, A is being observed from B; there is no notion of a stationary frame as such. Of course Einstein's verbal distinction can be temporarily applied to any frame (at any time) if desired.

$$_2t_A = {_1t_B}, \quad \text{(reciprocal events)} \tag{18}$$

$$_2t_A = \gamma {_2t_B} \quad \text{(Lorentz transform)} \tag{19}$$

$$_1t_B = \gamma {_1t_A} . \quad ( \text{``} \quad ) \tag{20}$$

The clocks in both frames have the same rate, so that between events 0 and 2 the B clock records a smaller elapsed time than that of A. (Eqn.19). Between 0 and 1, A records a smaller elapsed time than B (Eqn.20). There is therefore complete reciprocity between the frames and no feature enables one to be distinguished from the other.

Let us take it that frames A and B want to mutually find their separation so that they both get the same answer. Then, referring to Figure 5, they must each extend a measuring rod to the other at reciprocal times such as c and d or 1 and 2. They will use radar signals in practice, deriving distance from time; as Fock pointed out, rods are for Classical Mechanics, but light signals are for Relativity
We have:

$$_et_A = {_2t_A} = {_1t_B} = {_ft_B} \tag{21}$$

That is, (21) represents a simultaneity condition on the ends of the rod between events e and 2 in A, followed by reciprocal time equality between frames A and B and then simultaneity on the ends of the rod 1,f in B. We can therefore immediately write out distances/lengths:

$$_{e2}x_A = {_{f1}x_B} \tag{22}$$

On the other hand, the e, 2 distance in A could be compared with the c, 2 distance in B. We have the time relation $_2t_A = \gamma {_2t_B}$, that is, B finds a smaller time at event 2 than does A (eqn.19) and thus $_{e2}x_A = \gamma {_{c2}t_B}$. The events e and 2 are simultaneous in A; c, 2 are simultaneous in B, giving the respective distances or lengths between the frames. Frame B finds a smaller distance to event 2, where the two rods momentarily touch, but clearly it would be incorrect to argue that B's rod had contracted.

## 6.1 A Common View

I will briefly sketch what I believe is a fairly normal or received view[15]. Two standard clocks are synchronised at rest in the same location; one of them now moves around a closed circuit and back to the location of the "fixed" clock. On return, it is found that the moving clock has lost time according to the Lorentz transformation. Possible effects on the fixed clock are soon dismissed and the conclusion is that *the moving clock is the one affected by the motion*. The key operative is *affected* (stronger terms have been used). The notion is that the moving clock is caused to run slowly in the normally accepted sense (see p. 4). Often there is little further discussion about how it comes about that a moving clock should be affected. Rather it is straightforwardly accepted as a physical effect, in conformity with experiment and indeed what Special Relativity (known to be correct) says will happen.

So we have it that the "fixed" clock is unaffected, but the moving clock is affected. Returning to our example where we obtained equation (19); according to the view we are considering here it would be said that the clock in frame B was affected. The conclusion that clock B has been affected now makes it impossible to write equation 20. That is, we can no longer switch to frame B as fiducial, run another experiment, and get the required reciprocal result. Thus, any view which admits a physical slowing of a so called moving clock is inconsistent.

A common misconception sustained in these normal or received views is that the moving clock must return to the position of the "fixed clock" in order to demonstrate the time disparity; this is quite incorrect. In general (or take a familiar example such as the twin paradox) a simple straight path does all the work that is needed, since it immediately brings out the difference in clock readings between two inertial frames associated with a specific event set. The straight path also disenables spurious and incorrect discussion about acceleration accounting for time reading differences. Slowing of biological rhythms or the slowing of the 'clocks' of high speed particles are notions which add nothing to the discussion about clock rates - they merely add to a list of clock examples.

---

[15] Such views are of course contrary to mine. For example, quite current, is Johns, O. D. Amongst the vast literature of older texts one might consult Rindler, Fock and, say, Miller for a more historical survey. Harvey Brown takes the view that relativistic phenomena like length contraction and time dilation are, in the last analysis, the result of structural properties of the quantum theory of matter.

## 7  SR Summary

All of the standard results in Special Relativity remain the same, but their interpretation is changed in a most fundamental way. Good clocks are not affected by uniform motion; all clocks run at the same rate, recording elapsed time between event pairs in accordance with the Minkowski metric. In our treatment of length, we follow current metrological practice; length is defined in terms of the velocity of light by measuring time intervals. Of course, classical standard (say invar) measuring rods in all inertial frames conform to light signal/time measurements and no rod in any inertial frame becomes, so called, contracted.

Thorny questions in SR, which have come up again and again, find a straightforward resolution. Key questions centre on reciprocity; for example: why don't the clocks in both frames go slow? Neither clock goes slow – both clocks run at the same rate, there is complete symmetry between their respective readings. Most significantly, we distinguish between clocks which run slow (slowly) in the traditional sense (generally incorrectly invoked) and clocks which record a smaller time interval between a given pair of events while maintaining the standard rate. Thus for high speed atomic particles there is a smaller time interval between a given event pair than is found in the fiducial laboratory frame of reference. Finally, we have it that the underlying physics is changed: *clocks do not go slow and rods do not contract.*

# 8 General Relativity

When considering the behaviour of a clock in a gravitational field we speak of the gravitational redshift; again the term 'time dilation' (from SR) is also used[16]. This is the case we now turn to. Employing some standard bookwork[17] in a Schwarzschild spacetime, for a central mass M and a spherically symmetric universe, the line element is:

$$c^2 d\tau^2 = (1 - 2m/r)c^2 dt^2 - (1 - 2m/r)^{-1} dr^2 - r^2 d\vartheta^2 - r^2 \sin^2 \vartheta d\varphi^2, \quad (23)$$

where we have put $m = GM/c^2$. For a clock at a fixed point in space ($r, \vartheta, \phi,$ constant), we have:

$$d\tau = (1 - 2m/r)^{1/2} dt. \quad (24)$$

Fig.6 depicts the propagation of pulsed signals in a Schwarzschild spacetime. Signals are emitted at events 1 and 2 on radius $r_E$ and received at events 3 and 4 on radius $r_R$. An interval between such event pairs can be interpreted as the period of a high frequency electromagnetic signal – bringing our analysis into line with the more usual notion of the red shift. At the emitter and receiver respectively, we now have:

$$d\tau_E = (1 - 2m/r_E)^{1/2} dt_E \quad (25)$$

$$d\tau_R = (1 - 2m/r_R)^{1/2} dt_R. \quad (26)$$

We now take a stationary spacetime and a coordinate system such that $g_{\mu\nu,\gamma} = 0$; in such a coordinate system, the travel time for an electromagnetic signal is independent of the epoch. Thus, the interval between the reception of successive signals at the receiver is equal to the interval at the transmitter. In this way we establish that the coordinate clocks at the receiver and emitter run at the same rate. Synchrony according to Einstein can now be used to establish a coordinate time. We have:

$$_3t_R - {_1t_E} = {_4t_R} - {_2t_E},  \tag{27}$$

$$_2t_E - {_1t_E} = {_4t_R} - {_3t_R}.  \tag{28}$$

---

[16] Hafele and Keating, *Around the world atomic clocks: Observed Relativistic time gain,* relates to the S.R. and G.R. parts of this paper

[17] For example Misner, Thorn and Wheeler, *Gravitation* . There are many references under Schwarzshild, but I suggest Ch. 23, p593. Wald, *General Relativity,* Ch.6, pp 118 – 158, soon derives the Schwarzschild solution

We therefore set $dt_E = dt_R$, giving:

$$d\tau_R / d\tau_E = (1 - 2m/r_R)^{1/2} / (1 - 2m/r_E)^{1/2}.  \tag{29}$$

With $\tau$ proper time and $r_R > r_E$, $d\tau_R / d\tau_E > 1$.  (30)

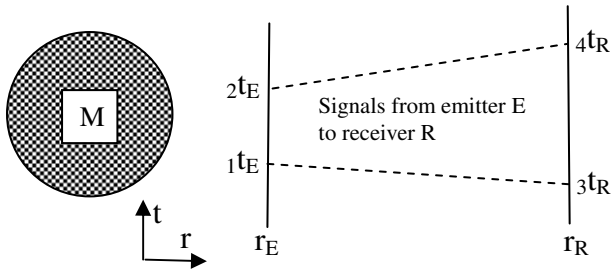

Figure 6. Redshift. A mass M in a Schwarzschild spacetime. Signals are emitted close to the mass and received at a more distant point

In (29), the proper time interval between signal pulses at the receiver is longer than the interval at the emitter; equivalently, the signal frequency at the receiver is lower than at the emitter. This is the gravitational red shift. In the SR case, as we have seen, the usual interpretation is that one clock runs slowly because of its motion. We would fall into a similar error now if we concluded that one clock runs more slowly because it is in a more intense gravitational field. We first note that the time intervals measured at the emitter and receiver are between different event pairs. We should not therefore reach conclusions about clock rates without appealing to a correct rating method.

We see from equation (29) that the red shift can be attributed to the different gravitational potentials at the transmitter and receiver and this is expressed in a proper time ratio. We know that good clocks in free fall obey SR and thus, running at the same rate, record proper time intervals between successive events at their instantaneous locations. The conditions required to measure the redshift proper times are therefore:

During the interval of signal emission, the associated clock is substantially at rest in free fall, at radius $r_E$.

During the interval of signal reception, the associated clock is substantially at rest in free fall at radius $r_R$

The situation now is that we have followed the correct logic by first establishing the equality of clock rates at different locations. We thus conclude: between an event pair at the receiver the associated clock records a longer time interval than the corresponding time interval recorded at the emitter - *both clocks running at the same rate*. Put another way: according to a given time standard, there is more time between an event pair at the receiver than there is between the corresponding events at the emitter.

In both SR and GR, we employ time intervals on standard clocks running at the standard rate. On what might be called the usual view in Special Relativity, clocks are said to be slowed on account of their motion, but no satisfactory explanation, or calculation, has ever been given. In General Relativity, a clock in a stronger gravitational field is said to run more slowly - again, a notion exists that the clock is in some way affected.

By considering what we mean by the rate of a clock and the amount of time between given event pairs, our conclusion is that clocks (accordingly rods) are not affected in Relativity Theory. Clocks do not go slow and rods do not contract.

**Acknowledgements**

I am permanently indebted to Professor C W Kilmister.